\documentclass[amsmath,amssymb,12pt,superscriptaddress,nofootinbib]{revtex4-1}

\usepackage{graphicx}
\usepackage{dcolumn}
\usepackage{bm}
\usepackage{color}
\usepackage{enumitem}
  \usepackage{tabularx}
   \newcolumntype{C}{>{\centering\arraybackslash}X}
   \newcolumntype{L}{>{\raggedright\arraybackslash}X}
   \newcolumntype{R}{>{\raggedleft\arraybackslash}X}
\setlistdepth{10}
\newcommand{\dd}{\mathrm{d}}

\newcommand{\del}{\partial}
\newcommand{\ee}{{\rm e}}

\definecolor{DarkBlue}{rgb}{0,0,0.7} 

\definecolor{DarkRed}{rgb}{0.65,0,0}

\begin{document}
\baselineskip5.5mm
%\lineskip5.5mm

%\if0

{\baselineskip0pt
\small
\leftline{\baselineskip16pt\sl\vbox to0pt{
%               \hbox{\it Division of Particle and Astrophysical Science, Nagoya University}
%%              \hbox{\it Instituto Superior T\'ecnico}
%%              \hbox{\it Nagoya University}
% %              \hbox{\it Department of Physics, Rikkyo University} 
%%             \hbox{\it Rikkyo University}
%%		\hbox{\it Yukawa Institute for Theoretical Physics, Waseda University} 
%%               \hbox{\it Advanced Research Institute for Science and Engineering, Waseda University} 
%%             \hbox{\it Waseda University}
                             \vss}}
\rightline{\baselineskip16pt\rm\vbox to20pt{
 		%\hbox{OCU-PHYS-xxx}
%            \hbox{AP-GR-xx}
\vspace{-1.5cm}
%            \hbox{RUP-18-15}
\vss}}
}

\author{Chul-Moon~Yoo}\email{yoo@gravity.phys.nagoya-u.ac.jp}
\affiliation{
\fontsize{11pt}{0pt}\selectfont
%Gravity and Particle Cosmology Group,
Division of Particle and Astrophysical Science,
Graduate School of Science, Nagoya University, 
Nagoya 464-8602, Japan
\vspace{1.5mm}
}

%\author{Vitor~Cardoso}\email{vitor.cardoso@ist.utl.pt}
%\affiliation{
%\fontsize{11pt}{0pt}\selectfont
%CENTRA, Departamento de F\'{\i}sica, Instituto Superior T\'ecnico, Universidade de Lisboa, Avenida~Rovisco Pais 1, 1049% Lisboa, Portugal
%\vspace{1.5mm}
%}
%\affiliation{~
%\fontsize{11pt}{0pt}\selectfont
%Perimeter Institute for Theoretical Physics, 31 Caroline Street North Waterloo, Ontario N2L 2Y5, Canada
%\vspace{1.5mm}
%}

\author{Taishi~Ikeda}\email{taishi.ikeda@tecnico.ulisboa.pt}
\affiliation{
\fontsize{11pt}{0pt}\selectfont
CENTRA, Departamento de F\'{\i}sica, Instituto Superior T\'ecnico, Universidade de Lisboa, Avenida~Rovisco Pais 1, 1049 Lisboa, Portugal
\vspace{1.5mm}
}

\author{Hirotada~Okawa}\email{h.okawa@aoni.waseda.jp}
\affiliation{
\fontsize{11pt}{0pt}\selectfont
Yukawa Institute for Theoretical Physics,
Kyoto University, Kyoto 606-8502, Japan
\vspace{1.5mm}
}

\affiliation{
\fontsize{11pt}{0pt}\selectfont
Advanced Research Institute for Science and Engineering, 
Waseda University, 3-4-1 Okubo, Shinjuku, 
Tokyo 171-8501, Japan
\vspace{1.5mm}
}

\vskip0.5cm
\title{Gravitational Collapse of a Massless Scalar Field \\in a Periodic Box}

\begin{abstract}
\baselineskip5mm 
\vskip0.5cm 
Gravitational collapse of a massless scalar field with 
the periodic boundary condition in a cubic box is reported. 
This system can be regarded as a lattice universe model. 
We construct the initial data for a Gaussian like profile of the scalar field 
taking the integrability condition associated with the periodic boundary condition into account. 
For a large initial amplitude, a black hole is formed after a certain period of time. 
While the scalar field spreads out in the whole region for a small initial amplitude. 
It is shown that the expansion law in a late time approaches to that of the radiation dominated universe and the matter dominated universe for the small and large initial amplitude cases, respectively. 
For the large initial amplitude case, 
the horizon is initially a past outer trapping horizon, whose area decreases with time, and after a certain period of time, it turns to a future outer trapping horizon with the increasing area. 
\end{abstract}

\maketitle

\section{Introduction}
\label{intro}
When we consider gravitational collapse, 
the most common boundary condition is the asymptotically flat 
boundary condition. 
Obviously, the asymptotically flat boundary condition is relevant for stellar collapse. 
Nonetheless, when we consider collapse of a cosmological object, 
we may not neglect the effect of the expansion of the universe. 
For instance, for simulation of a primordial black hole~\cite{1967SvA....10..602Z,Hawking:1971ei}, 
the black hole horizon scale is typically comparable to 
the Hubble horizon scale. 
Here, we report a numerical experiment of gravitational collapse in an expanding background. 

In this paper, we consider a massless scalar field in a cubic box with 
the periodic boundary condition for each pair of opposite boundary surfaces. 
This system can be regarded as a lattice universe model in which the same cubic region is spatially repeated. 
A representative example of the lattice universe models is the black hole lattice universe, 
which is intensively studied in recent years~\cite{RevModPhys.29.432,Clifton:2009jw,Clifton:2012qh,Bentivegna:2012ei,Yoo:2012jz,Bentivegna:2012ei,Bruneton:2012cg,Bruneton:2012ru,Bentivegna:2013xna,Yoo:2013yea,Bentivegna:2013jta,Clifton:2013jpa,Korzynski:2013tea,Clifton:2014lha,Yoo:2014boa,Bentivegna:2016fls,Clifton:2017hvg}(see a recent review~\cite{Bentivegna:2018koh} and references therein).% 
\footnote{
\baselineskip5mm
One notable example of a lattice universe model with matter fields is 
analyzed in Ref.~\cite{Clifton:2017hvg}, where a persistent black hole lattice universe model in bouncing cosmology is 
discussed. 
A scalar field is introduced to realize the bouncing cosmology with violation of the null energy condition due to the scalar field. 
In our case, we consider a massless scalar field, and there is no energy condition violation.  
}
In practice, we simulate the only octant region of the cubic box imposing reflecting boundary 
condition on each boundary surface. 
That is, we focus on symmetric configurations where the reflecting boundary 
condition on each boundary surface of the octant region is realized 
as a consequence of the dynamics with the periodic boundary condition 
for a cubic box.% 
\footnote{\baselineskip5mm
We note that our boundary condition is essentially different from the settings 
of a scalar field enclosed in spherical cavity discussed in, e.g., Refs.~\cite{Bizon:2011gg,Maliborski:2012gx}. 
}
The initial scalar field configuration is set to be a spherically symmetric Gaussian like profile. 
Then, we report how the dynamics of the system changes 
depending on the initial amplitude of the scalar field profile focusing on 
black hole formation and expansion law of the lattice universe. 

In Sec.~\ref{sec-2}, settings and initial data construction are shown. 
In Sec.~\ref{sec-3}, we show results on a sequence of initial data sets. 
The time evolution for large amplitude and small amplitude cases are reported in Sec.~\ref{sec-4}. 
We use the geometrized units in which both
the speed of light and Newton's gravitational constant are 
one. 
Greek indices run from 0 to 3 and Latin indices run from 1 to 3. 

\section{Settings and initial data construction}
\label{sec-2}

\subsection{EoM for a scalar field}

We consider the massless scalar field $\phi$, whose equations of motion are given by 
\begin{equation}
\nabla^\mu \nabla_\mu \phi=0. 
\end{equation}
Introducing the momentum 
\begin{equation}
\Pi:=-n^\mu \nabla_\mu \phi, 
\end{equation}
we can rewrite the field equations in the form of the 3+1 decomposition 
as follows:
\begin{eqnarray}
&&\left(\del_t-\beta^i\del_i\right)\phi=-\alpha\Pi, \\
&&\left(\del_t-\beta^i\del_i\right)\Pi=-\alpha\triangle\phi-\gamma^{ij}\del_i\alpha \del_j\phi
+\alpha K \Pi,  
\end{eqnarray}
where $n^\mu$, $\alpha$, $\beta^i$, $\gamma^{ij}$, $K$ and $\triangle$ are 
the unit normal vector to a time slice, lapse function, shift vector, inverse spatial metric, 
the trace of the extrinsic curvature $K_{ij}$ and the Laplacian with respect to the spatial metric $\gamma_{\mu\nu}$,
respectively.

\subsection{Stress energy}

The stress energy tensor
\begin{equation}
T_{\mu\nu}=\nabla_\mu\phi \nabla_\nu \phi -g_{\mu\nu}\left(\frac{1}{2}\nabla^\rho\phi\nabla_\rho\phi\right)
\end{equation}
can be also expressed in the form of the 3+1 decomposition as follows: 
\begin{eqnarray}
\rho&:=&n_\mu n_\nu T^{\mu\nu}=\frac{1}{2}\Pi^2+\frac{1}{2}\ee^{-4\psi}\tilde \gamma^{ij}\del_i\phi\del_j\phi, \\
j_i&:=&-\gamma_{i\mu}n_\nu T^{\mu\nu}=\Pi\del_i\phi, \\
S_{ij}&:=&\gamma_{i\mu}\gamma_{j\nu}T^{\mu\nu}=\del_i\phi \del_j\phi
-\frac{1}{2}\tilde \gamma_{ij}\tilde \gamma^{kl}
\del_k\phi \del _l \phi+\frac{1}{2}\ee^{4\psi}\tilde \gamma_{ij}\Pi^2, 
\end{eqnarray}
where $\gamma_{ij}=\ee^{4\psi}\tilde \gamma_{ij}$.

\subsection{Constraint equations}

Constraint equations are given by the followings:
\begin{eqnarray}
&\mathcal{R}+K^2-K_{ij}K^{ij}&=16\pi \rho, \\
&D_jK^{ji}-D^iK&=8\pi j_i. 
\end{eqnarray}
Under the conventional decomposition given by 
\begin{equation}
\dd l^2=\gamma_{ij}\dd x^i \dd x^j
=\Psi^4\tilde \gamma_{ij}\dd x^i \dd x^j
=\ee^{4\psi}\tilde \gamma_{ij}\dd x^i \dd x^j
\end{equation}
and
\begin{equation}
K_{ij}=\ee^{4\psi}\tilde A_{ij}+\frac{1}{3}K\gamma_{ij}, 
\end{equation}
the constraint equations are rewritten as 
\begin{eqnarray}
0&=&\tilde D_i\tilde D^i\Psi-\frac{1}{8}\tilde {\mathcal R}\Psi+
\left(2\pi\rho+\frac{1}{8}\tilde A_{ij}\tilde A^{ij}-\frac{1}{12}K^2\right)\Psi^5,  \\
0&=&\tilde D_j\tilde A^{ij}+6\tilde A^{ij}\tilde D_j\ln \Psi-\frac{2}{3}\tilde D^i K-8\pi\Psi^4j^i, 
\end{eqnarray}
where $\tilde D_i$ and $\tilde {\mathcal R}$ are the covariant derivative and 
Ricci scalar with respect to the metric $\tilde \gamma$. 
We require that the initial data should satisfy the following conditions: 
\begin{eqnarray}
\Pi&=&0, \\
K&=&{\rm const.}, \\
\tilde \gamma_{ij}&=&f_{ij}, \\
\tilde A_{ij}&=&0, 
\end{eqnarray}
where $f_{ij}$ is a flat metric. 
Then, the momentum constraints are trivially satisfied with $j_i=0$. 
The Hamiltonian constraint reduces to 
\begin{equation}
\hat \triangle \psi +\delta^{ij}\del_i\psi\del_j\psi-\frac{1}{12}K^2\ee^{4\psi}=-2\pi\rho\ee^{4\psi}
=-\pi f^{ij}\del_i\phi\del_j\phi, 
\label{eq:ham}
\end{equation}
where $\hat \triangle$ is the flat Laplacian 
with respect to $f_{ij}$.

\subsection{Scale-up coordinate}

\begin{figure}[htbp]
\begin{center}
\includegraphics[scale=0.7]{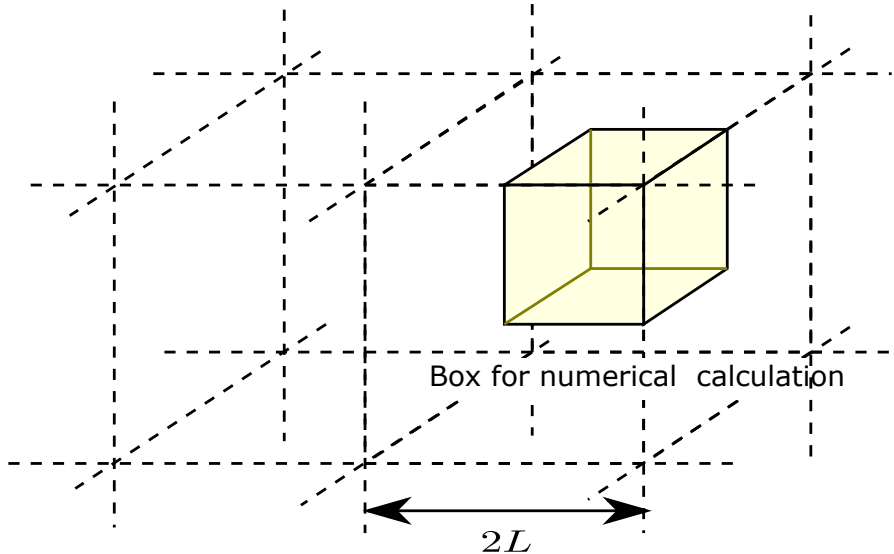}
\caption{Domain of numerical simulation. 
}
\label{fig:cube}
\end{center}
\end{figure}

The system is located inside the cubic region 
specified by $-L\leq X^a \leq L~(a=1,2,3)$, 
with the isotropic Cartesian coordinates $X^a$, 
for which the flat metric components are given by the unit matrix $\delta_{ab}$. 
Hereafter, for convenience, we 
use the indices $a,b,\cdots$ to specify each component of an object 
not applying the summation rule to them. 
Assuming reflection symmetry 
on each surface of $X^a=0$, we perform numerical simulations  
in 1/8 domain of cubic region given by 
$0\leq X^a \leq L$(see Fig.\ref{fig:cube}). 
%%%%%%%%%%%%%%%%%%%%%%%%%%%<<start figure>>%%%%%%%%%%%%%%%%%%%%%%%%%%
%%%%%%%%%%%%%%%%%%%%%%%%%%%%<<end figure>>%%%%%%%%%%%%%%%%%%%%%%%%%%%

In order to see the horizon formation, 
it is better to have finer grid spacing in the central region 
of gravitational collapse. 
For this purpose, we introduce the ``scale-up" grids based on 
a coordinate system $x^a$ defined as follows:
\begin{equation}
X^a=F(x^a) ~~{\rm for~each~}a 
\end{equation}
with
\begin{equation}
F(x):=x-\frac{\eta}{1+\eta}\frac{L}{\pi}\sin\left(\frac{\pi}{L}x\right), 
\label{eq:funcf}
\end{equation}
where $\eta$ is a parameter which specifies the degree of the scale-up. 
The function $F(x)$ has the following properties: 
\begin{itemize}
\item
$F(x^a)$ is compatible with the reflection boundary condition 
on each face. 

\item
$x^a=0$ for $X^a=0$, and $x^a=L$ for $X^a=L$. 

\item
$\left.\frac{\dd X^a}{\dd x^a}\right|_{x^a=L}
\big/\left.\frac{\dd X^a}{\dd x^a}\right|_{x^a=0}=1+2\eta$

\end{itemize}
Using the coordinates $x^a$, non-vanishing flat metric components are 
given by 
\begin{eqnarray}
f_{aa}%={F'}^2_a:
=\left(\frac{\dd F(x^a)}{\dd x^a}\right)^2. 
\end{eqnarray}

%%%%%%%%%%%%%%%%%%%%%%%%%%%%%%%%%%%%%%%%%%%%%%%%%%%%%%%%%%%%%%%%
\subsection{Initial data}
\label{sec:integrability}
%%%%%%%%%%%%%%%%%%%%%%%%%%%%%%%%%%%%%%%%%%%%%%%%%%%%%%%%%%%%%%%%

We set the profile of the initial scalar field as follows:
\begin{equation}
\phi_0=A\exp\left[-\frac{R^2}{2\sigma^2}\right]
W(R;R_{\rm W},L), 
\label{eq:initial_scalar}
\end{equation}
where $R=\sqrt{X^2+Y^2+Z^2}$ and $W(R;R_{\rm W},L)$ is give by 
\begin{equation}
W(R;R_{\rm W},L)=
\left\{
\begin{array}{ll}
1&{\rm for}~0\leq R \leq R_{\rm W} \\
1-\frac{\left(\left(R_{\rm W}-L\right)^6-\left(L-R\right)^6\right)^6}{\left(R_{\rm W}-L\right)^{36}}&{\rm for}~R_{\rm W} \leq R \leq L 
\\
0&{\rm for}~L \leq R 
\end{array}\right..
\end{equation}
We have introduced the function $W$ to make the Gaussian tail gradually disappear toward the boundary. 

In order to solve Eq.~\eqref{eq:ham}, 
we need to impose the following integrability condition:
\begin{eqnarray}
0&=&\int_{\rm box}\sqrt{f}\dd ^3 x
\left(\triangle\psi-\frac{1}{12}K^2\ee^{4\psi}+(\del \psi)^2
+\pi(\del \phi)^2\right)
\cr
&\Leftrightarrow&K^2
=12
\frac{\int_{\rm box}\sqrt{f}\dd ^3 x \left((\del\psi)^2+\pi(\del\phi)^2\right)}
{\int_{\rm box}\sqrt{f}\dd ^3x \ee^{4\psi}}, 
\end{eqnarray}
where $f$ is the determinant of $f_{ij}$ and the integral of the Laplacian part vanishes because of 
the periodic boundary condition. 
We numerically solve the elliptic differential equation \eqref{eq:ham} 
by using iterative method imposing the integrability condition 
at each step of the iteration. 
The length scale of the coordinates is fixed by 
imposing $\psi(L,L,L)=0$.

%%%%%%%%%%%%%%%%%%%%%%%%%%%%%%%%%%%%%%%%%%%%%%%%%%%%%%%%%%%%%%%%
\section{Results of Initial Data}
\label{sec-3}
%%%%%%%%%%%%%%%%%%%%%%%%%%%%%%%%%%%%%%%%%%%%%%%%%%%%%%%%%%%%%%%%

\begin{figure}[htbp]
\begin{center}
\includegraphics[scale=1.2]{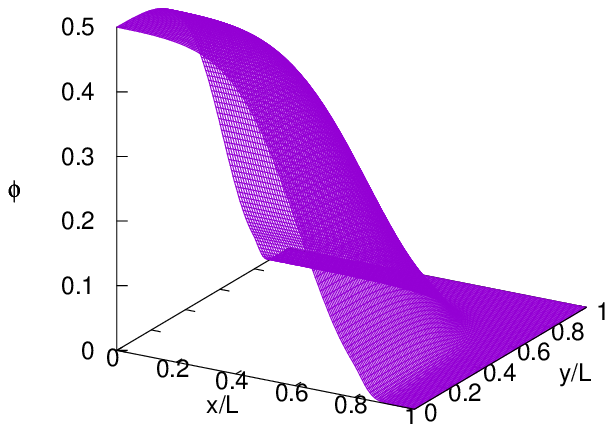}
\includegraphics[scale=1.2]{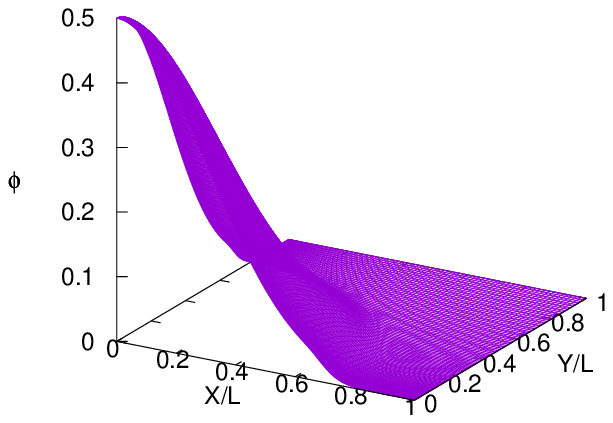}
\includegraphics[scale=1.2]{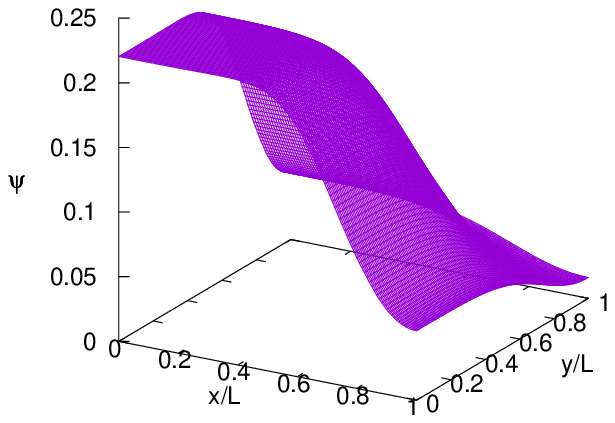}
\includegraphics[scale=1.2]{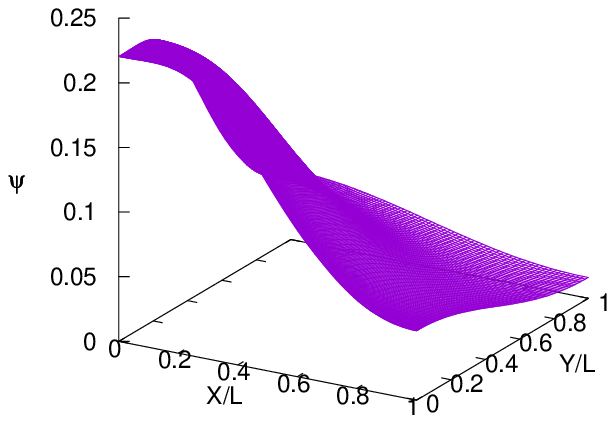}
\caption{Initial profiles of $\phi$ and $\psi$ for $A=0.5$. 
The left two figures are plotted with the scale-up coordinates $x$ and $y$, while the isotropic coordinates $X$ and $Y$ are used in the right two figures. 
}
\label{fig:A05}
\end{center}
\end{figure}

\begin{figure}[htbp]
\begin{center}
\includegraphics[scale=1.2]{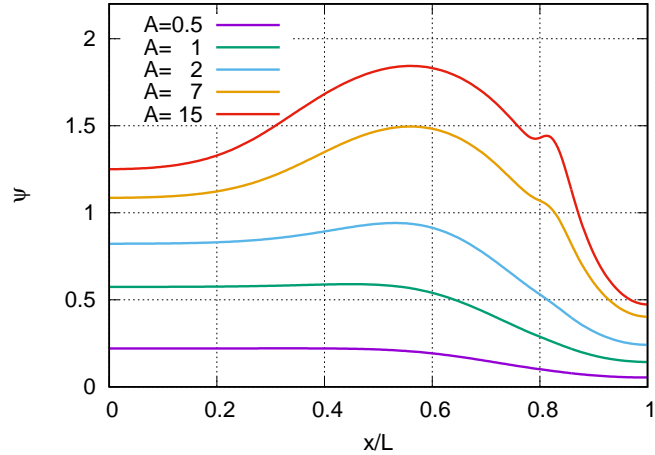}
\caption{The value of $\psi$ at the initial time on $x$-axis
is depicted as a function of $x$ for each value of the amplitude $A$. 
}
\label{fig:psi_Adep}
\end{center}
\end{figure}

\begin{figure}[htbp]
\begin{center}
\includegraphics[scale=1.2]{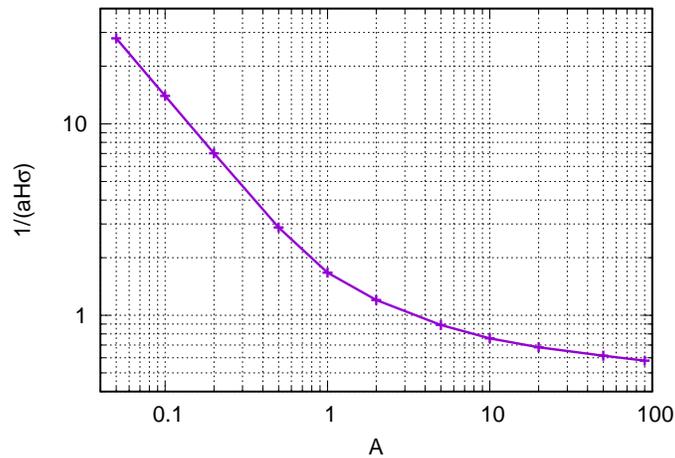}
\caption{The ratio between the effective Hubble length $1/H=-3/K$ and 
the proper width $a\sigma$ of the scalar field profile  
as a function of $A$. 
}
\label{fig:AHi}
\end{center}
\end{figure}

We set $R_W=0.6L$, $\sigma=0.3L$ and $\eta=5$ for the results in this section. 
First, we show a typical initial data profile on the $x$-$y$ plane for $A=0.5$ in 
Fig.~\ref{fig:A05}. 
%%%%%%%%%%%%%%%%%%%%%%%%%%%<<start figure>>%%%%%%%%%%%%%%%%%%%%%%%%%%
%%%%%%%%%%%%%%%%%%%%%%%%%%%%<<end figure>>%%%%%%%%%%%%%%%%%%%%%%%%%%%
The $A$ dependence of the profile of the initial $\psi$ on the $x$-axis is shown in Fig.~\ref{fig:psi_Adep}. 
%%%%%%%%%%%%%%%%%%%%%%%%%%%<<start figure>>%%%%%%%%%%%%%%%%%%%%%%%%%%
%%%%%%%%%%%%%%%%%%%%%%%%%%%%<<end figure>>%%%%%%%%%%%%%%%%%%%%%%%%%%%
The small bump around $x=0.8M$ for $A=7$ and $15$ is 
caused by the effect of the function $W$ introduced in Eq.~\eqref{eq:initial_scalar}.  
As is shown in Sec.~\ref{sec:integrability}, 
the value of $K$ is determined so that the integrability condition 
can be satisfied for each value of $A$. 
Letting $\mathcal S$ denote the area of the surface of the cubic region 
with the edge length $2L$, we can define the scale factor $a$ by $\sqrt{\mathcal S/(24L^2)}$.  
The ratio between the effective Hubble length $1/H:=-3/K$ and 
the proper width $a\sigma$ of the scalar field profile  
is depicted as a function of $A$ in Fig.~\ref{fig:AHi}. 
As is shown in Fig.~\ref{fig:AHi}, the scale of the inhomogeneity is 
comparable or larger than the Hubble scale for $A\gtrsim1$. 
%%%%%%%%%%%%%%%%%%%%%%%%%%%<<start figure>>%%%%%%%%%%%%%%%%%%%%%%%%%%
%%%%%%%%%%%%%%%%%%%%%%%%%%%%<<end figure>>%%%%%%%%%%%%%%%%%%%%%%%%%%%

%%%%%%%%%%%%%%%%%%%%%%%%%%%%%%%%%%%%%%%%%%%%%%%%%%%%%%%%%%%%%%%%
\section{Time Evolution}
\label{sec-4}
%%%%%%%%%%%%%%%%%%%%%%%%%%%%%%%%%%%%%%%%%%%%%%%%%%%%%%%%%%%%%%%%

%We show results of two typical cases. 
%One is the case for $A=0.1$ with $\eta=0$, and the other is for $A=10$ with $\eta=5$. 
For the simulation of the time evolution, we use the 4th order Runge-Kutta method with 
the so-called BSSN formalism~\cite{Shibata:1995we,Baumgarte:1998te}. 
We use the same gauge condition as in Ref.~\cite{Yoo:2013yea}, and take 80 - 120 number of grids for each side. 
The grid spacing is uni-grid with the Cartesian coordinates for $A=0.1$ and 
the scale-up coordinates introduced in Sec.~\ref{sec-2} for $A=3$, $5$ and $10$. 
Excision of a black hole interior region\cite{Alcubierre:2000yz} is performed when a black hole is 
formed. 

%%%%%%%%%%%%%%%%%%%%%%%%%%%%%%%%%%%%%%%%%%%%%%%%%%%%%%%%%%%%%%%%
\subsection{$A=0.1$: scalar field diffusion}
\begin{figure}[htbp]
\begin{center}
\includegraphics[scale=0.82]{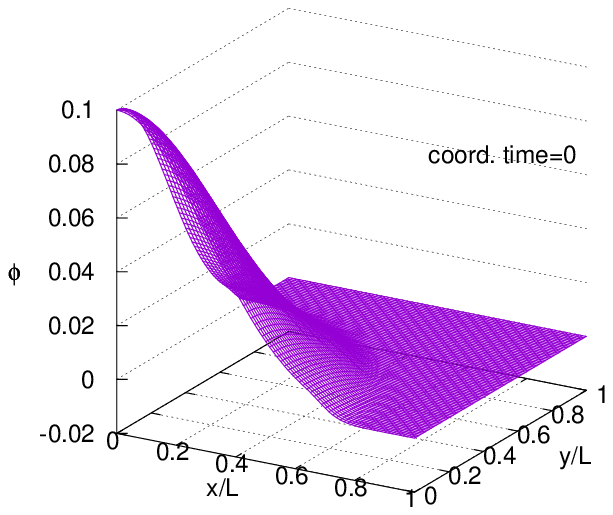}
\includegraphics[scale=0.82]{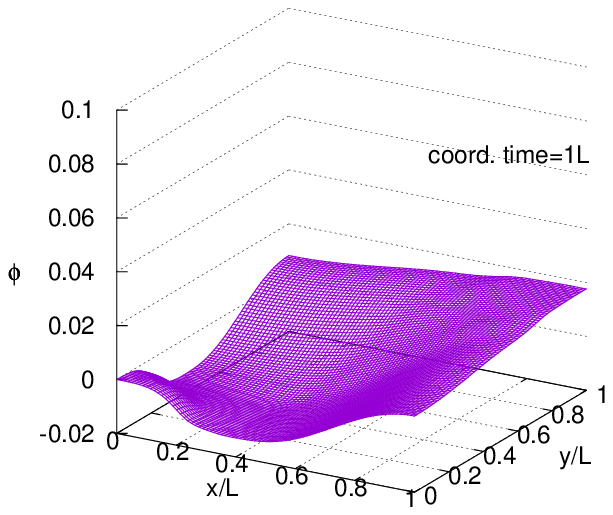}
\includegraphics[scale=0.82]{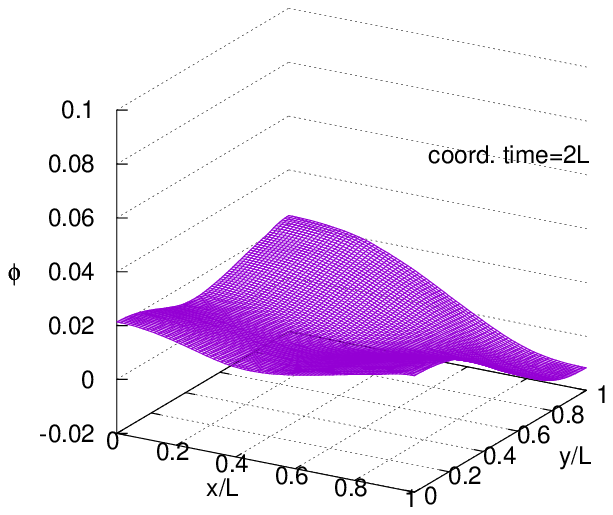}
\includegraphics[scale=0.82]{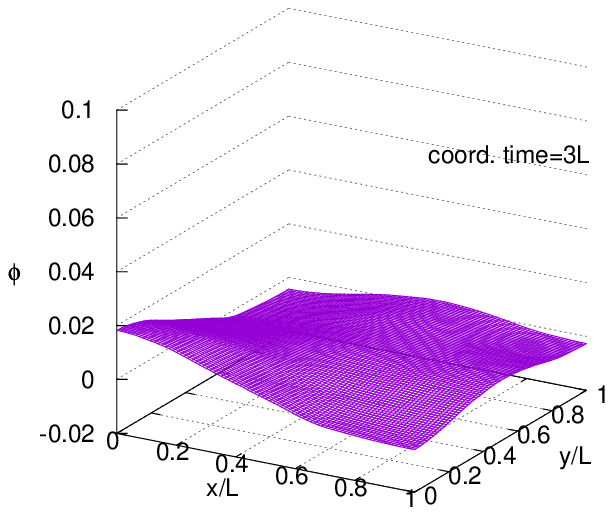}
\includegraphics[scale=0.82]{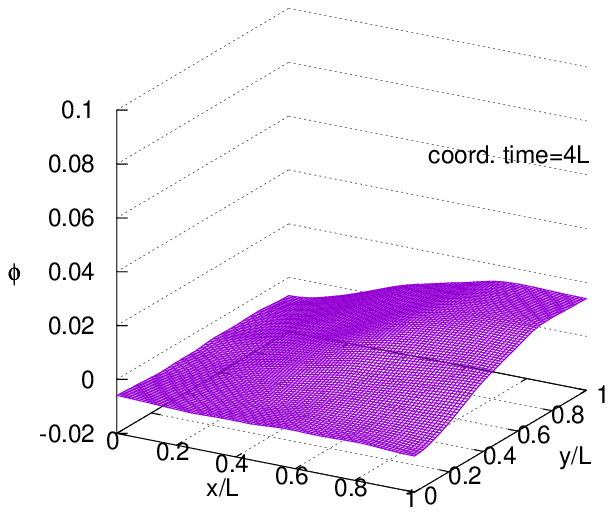}
\includegraphics[scale=0.82]{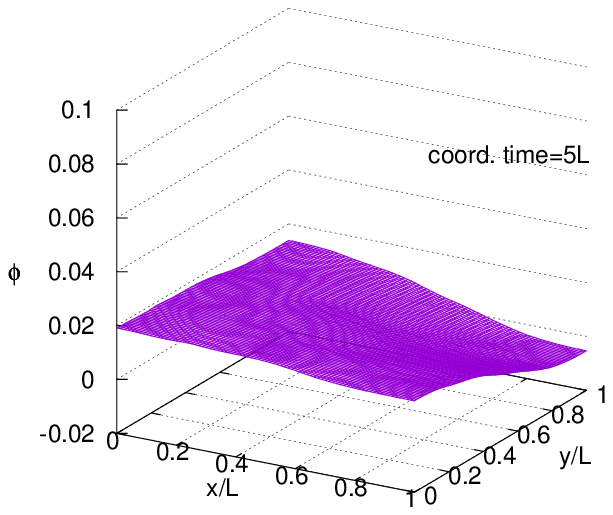}
\caption{Snapshots of the configuration of the scalar field $\phi$ on $z=0$ surface. 
}
\label{fig:phis}
\end{center}
\end{figure}

\begin{figure}[htbp]
\begin{center}
\includegraphics[scale=1.1]{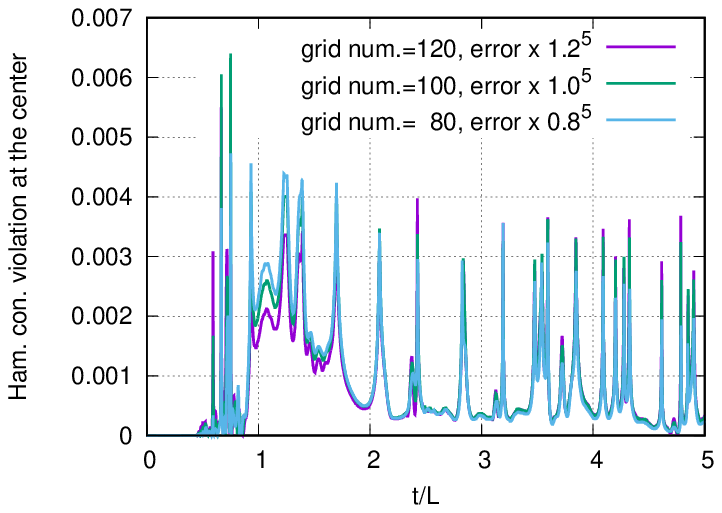}
\includegraphics[scale=1.1]{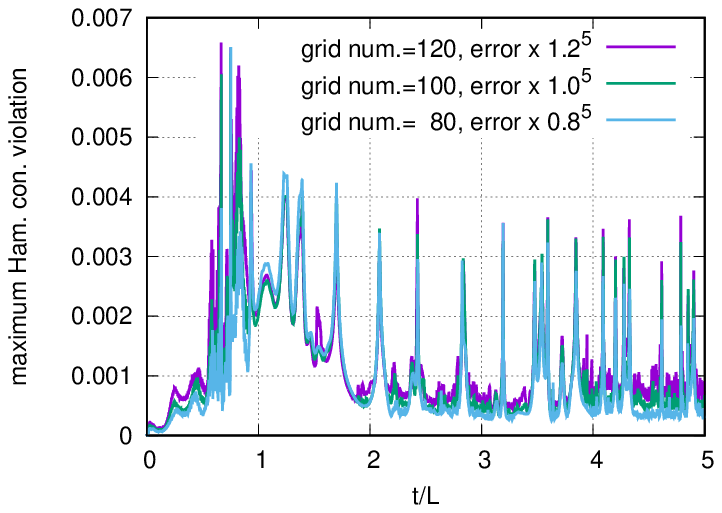}
\caption{The value of the Hamiltonian constraint violation at the center(left) and the max norm of the Hamiltonian constraint violation(right) are depicted as functions of 
the coordinate time $t$ for each resolution. 
A numerical factor proportional to the 5th power of the grid interval is multiplied to the value of the 
Hamiltonian constraint violation. 
}
\label{fig:const_conv}
\end{center}
\end{figure}

\begin{figure}[htbp]
\begin{center}
\includegraphics[scale=1.2]{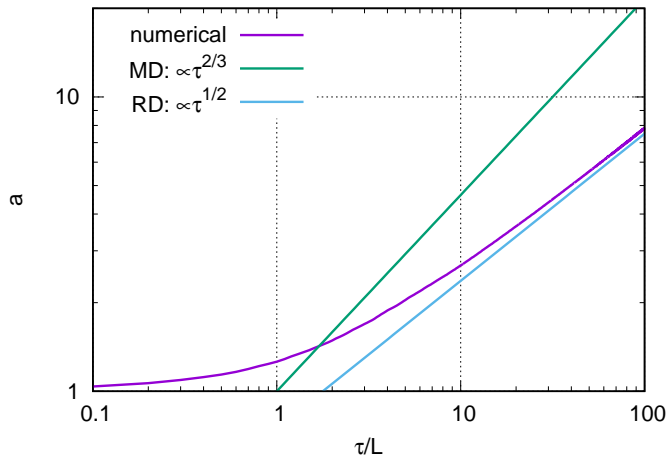}
\caption{The scale factor $a$ as a function of the proper time $\tau$ for $A=0.1$ with $\eta=0$. 
}
\label{fig:atau1}
\end{center}
\end{figure}

%%%%%%%%%%%%%%%%%%%%%%%%%%%%%%%%%%%%%%%%%%%%%%%%%%%%%%%%%%%%%%%%
For $A=0.1$, as is shown in Fig.~\ref{fig:phis}, 
the scalar field spreads into the whole box region and continue to oscillate. 
%%%%%%%%%%%%%%%%%%%%%%%%%%%<<start figure>>%%%%%%%%%%%%%%%%%%%%%%%%%%
%%%%%%%%%%%%%%%%%%%%%%%%%%%%<<end figure>>%%%%%%%%%%%%%%%%%%%%%%%%%%%
The oscillation amplitude decays with time due to the expansion of the whole system. 
The convergence of the Hamiltonian constraint violation at the center is shown in Fig.~\ref{fig:const_conv}. 
%%%%%%%%%%%%%%%%%%%%%%%%%%%<<start figure>>%%%%%%%%%%%%%%%%%%%%%%%%%%
%%%%%%%%%%%%%%%%%%%%%%%%%%%%<<end figure>>%%%%%%%%%%%%%%%%%%%%%%%%%%%
%
In order to see the expansion law of the system, let us consider the time evolution of the 
scale factor defined by the surface area of the box region. 
In order to avoid a highly distorted time slices for the calculation 
of the surface area, similarly to Refs.~\cite{Yoo:2013yea,Yoo:2014boa},  we calculate the surface area $\mathcal S$ on the constant proper time slice for the set of observers fixed at grid points. 
Thus, the area $\mathcal S$ is given as a function of the proper time $\tau$. 
The scale factor is defined as $a(\tau):=\sqrt{\mathcal S(\tau)/(24L^2)}$ as in Sec.~\ref{sec-3}. 
We note that the scale factor can be also defined by using the volume of the whole box, 
but we adopt the definition given by the surface area so that it can be directly 
compared with the case with black hole formation where the volume inside the black hole becomes 
highly non-trivial. 
We also emphasize that the expansion law defined by the scale factor is different from the local 
expansion law given at each point since the scale factor is defined by the integrated area of the boundary surface. 
The value of the scale factor $a$ is depicted as a function of the proper time $\tau$ in Fig.~\ref{fig:atau1}. 
%%%%%%%%%%%%%%%%%%%%%%%%%%%<<start figure>>%%%%%%%%%%%%%%%%%%%%%%%%%%
%%%%%%%%%%%%%%%%%%%%%%%%%%%%<<end figure>>%%%%%%%%%%%%%%%%%%%%%%%%%%%
As is shown in Fig.~\ref{fig:atau1}, the time evolution of the scale factor asymptotically approaches to that 
of the radiation dominated universe, that is $a\propto \tau^{1/2}$. 
This is the expected result because the oscillating massless scalar field behaves 
like a radiation fluid component on average in a homogeneous and isotropic universe(see also Ref.~\cite{Ikeda:2015hqa} for the case of gravitational waves). 
We note that for comparison, in the case of the homogeneous distribution of the massless scalar field, 
we obtain $a\propto \tau^{1/3}$. 

%%%%%%%%%%%%%%%%%%%%%%%%%%%%%%%%%%%%%%%%%%%%%%%%%%%%%%%%%%%%%%%%
\subsection{$A=3$, $5$ and $10$: black hole formation}
\begin{figure}[htbp]
\begin{center}
\includegraphics[scale=0.85]{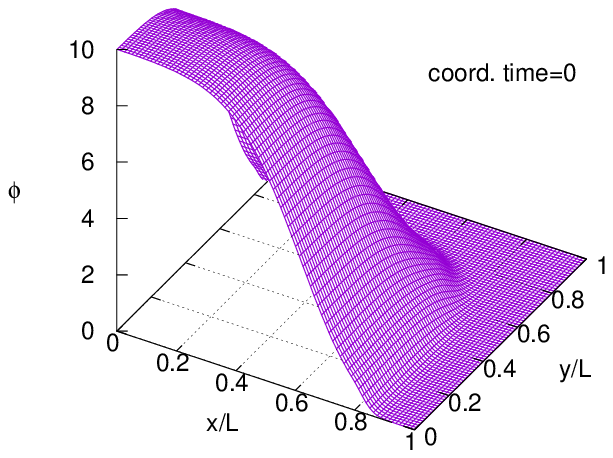}
\includegraphics[scale=0.85]{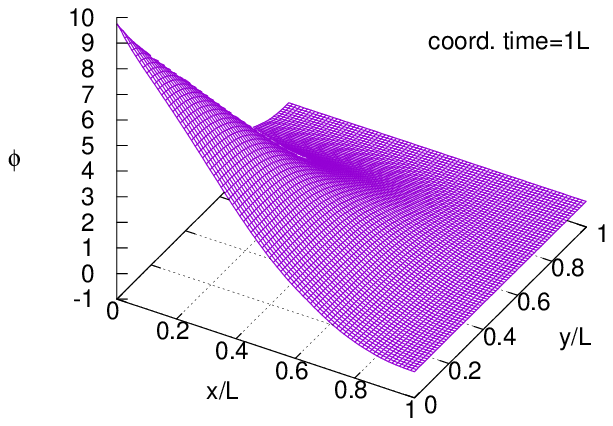}
\includegraphics[scale=0.85]{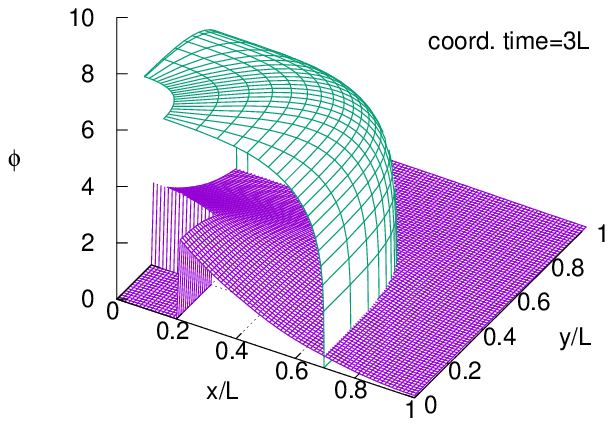}
\caption{The profiles of $\phi$ on the $z=0$ surface at 
the coordinate time $t=0$, $1L$ and $3L$ for $A=10$. 
The $z$ axis for the horizon figure is identified with the $\phi$ axis with some appropriate scaling. 
The central region is excised for $t=3L$. 
}
\label{fig:phis_A10}
\end{center}
\end{figure}

\begin{figure}[htbp]
\begin{center}
\includegraphics[scale=0.85]{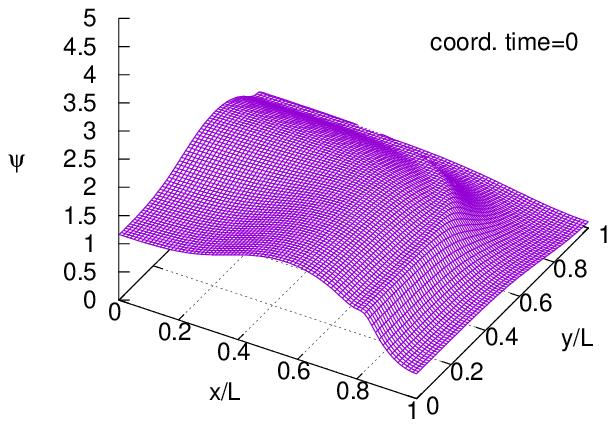}
\includegraphics[scale=0.85]{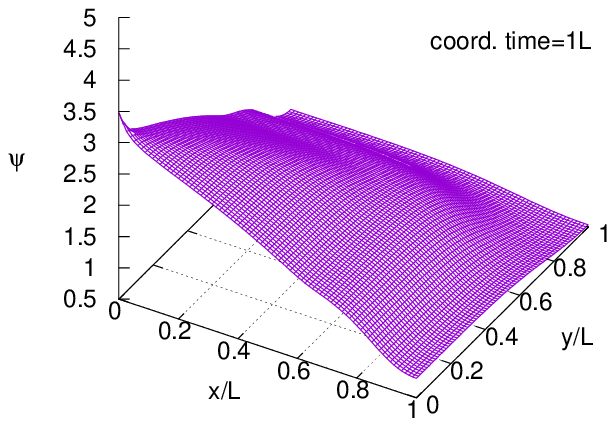}
\includegraphics[scale=0.85]{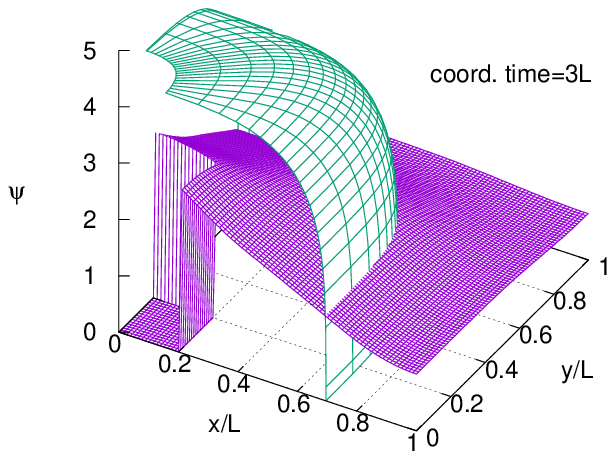}
\caption{The profiles of $\psi$ on the $z=0$ surface at 
the coordinate time $t=0$, $1L$ and $3L$ for $A=10$. 
The $z$ axis for the horizon figure is identified with the $\psi$ axis with some appropriate scaling. 
The central region is excised for $t=3L$. 
}
\label{fig:psis_A10}
\end{center}
\end{figure}

\begin{figure}[htbp]
\begin{center}
\includegraphics[scale=1.05]{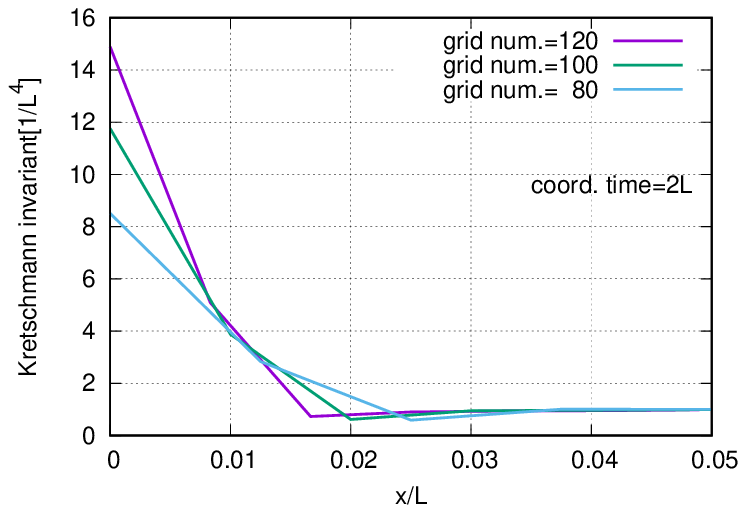}
\includegraphics[scale=1.05]{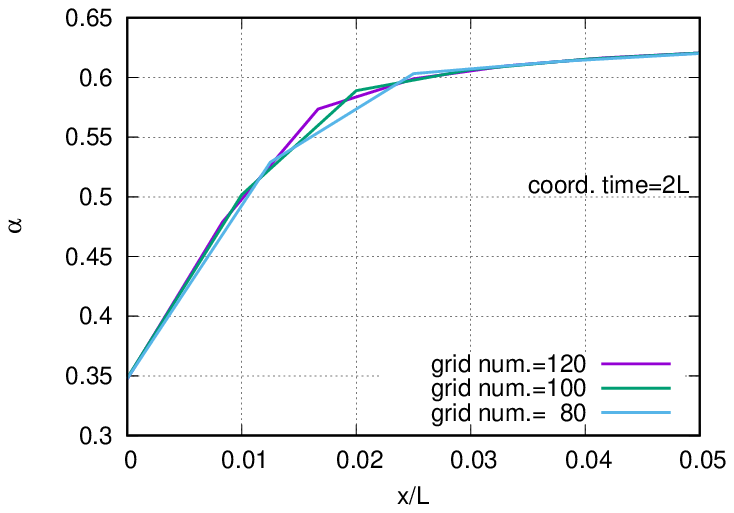}
\caption{The values of he Kretschmann invariant $R^{\mu\nu\rho\lambda}R_{\mu\nu\rho\lambda}$(left) and 
the lapse function $\alpha$(right) on the $x$--axis are depicted as functions of $x$ 
at the coordinate time $t=2L$ for $A=10$. 
}
\label{fig:cusp}
\end{center}
\end{figure}

\begin{figure}[htbp]
\begin{center}
\includegraphics[scale=1.2]{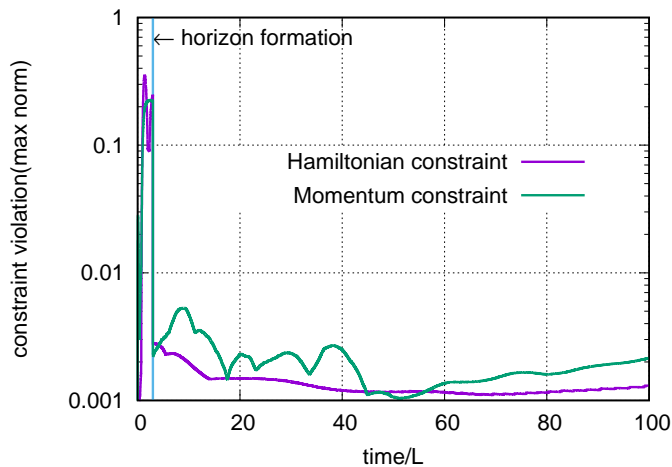}
\caption{Violations of the Hamiltonian constraint and the momentum constraint are depicted 
as functions of the coordinate time for $A=10$. 
The number of grids is set as 120.  
The constraint violations are evaluated by the maximum values of 
the normalized constraint violations calculated at all grid points 
except points inside the horizon when it exists. 
}
\label{fig:check}
\end{center}
\end{figure}

%%%%%%%%%%%%%%%%%%%%%%%%%%%%%%%%%%%%%%%%%%%%%%%%%%%%%%%%%%%%%%%%
For the $A=10$ case, 
a black hole is formed at around the coordinate time $t=2L$. 
We show the profiles of $\phi$ and $\psi$ on the $z=0$ surface at 
the coordinate time $t=0$, $1L$ and $3L$ in Figs.~\ref{fig:phis_A10} and \ref{fig:psis_A10}. 
At $t=3L$, a horizon exists, and the horizon figure is superposed in the figures at $t=3L$. 
The $z$ axis for the horizon figure is identified with the $\phi$ and $\psi$ axes with some appropriate scaling 
in Figs.~\ref{fig:phis_A10} and \ref{fig:psis_A10}, respectively. 
The central region is excised for $t=3L$. 
%%%%%%%%%%%%%%%%%%%%%%%%%%%<<start figure>>%%%%%%%%%%%%%%%%%%%%%%%%%%
%%%%%%%%%%%%%%%%%%%%%%%%%%%%<<end figure>>%%%%%%%%%%%%%%%%%%%%%%%%%%%
%%%%%%%%%%%%%%%%%%%%%%%%%%%<<start figure>>%%%%%%%%%%%%%%%%%%%%%%%%%%
%%%%%%%%%%%%%%%%%%%%%%%%%%%%<<end figure>>%%%%%%%%%%%%%%%%%%%%%%%%%%%
It should be mentioned that the behavior of $\psi$ and $\phi$ near the center is not regular even before 
the horizon formation. 
Actually, we find that constraints are violated near the center. 
We have checked that this behavior is qualitatively not sensitive to the resolution. 
One may understand the reason of this cusp at the center from Fig.~\ref{fig:cusp}. 
%%%%%%%%%%%%%%%%%%%%%%%%%%%<<start figure>>%%%%%%%%%%%%%%%%%%%%%%%%%%
%%%%%%%%%%%%%%%%%%%%%%%%%%%%<<end figure>>%%%%%%%%%%%%%%%%%%%%%%%%%%%
In Fig.~\ref{fig:cusp}, the values of the Kretschmann invariant $R^{\mu\nu\rho\lambda}R_{\mu\nu\rho\lambda}$(left) and 
the lapse function $\alpha$(right) on the $x$--axis are depicted as functions of $x$ at the coordinate time $t=2L$. 
It is found that the value of the Kretschmann invariant increases at the center and the value of the lapse function 
$\alpha$ decreases, so that the time evolution at the center gradually slows down. 
Our resolution near the center is not enough to accurately resolve this behavior and the constraints are significantly violated near the center. 
Nevertheless,
since it is well inside the radius of the horizon which will be formed, 
this region seems not to be causally connected with the spacetime domain outside the horizon in the future. 
That is, we may expect that the singular behavior does not affect the evolution of the system outside the horizon 
after its formation. 
In fact, constraint violation is well suppressed outside the horizon, 
and we ignore this singular behavior in the center(see Fig.~\ref{fig:check}). 
%%%%%%%%%%%%%%%%%%%%%%%%%%%<<start figure>>%%%%%%%%%%%%%%%%%%%%%%%%%%
%%%%%%%%%%%%%%%%%%%%%%%%%%%%<<end figure>>%%%%%%%%%%%%%%%%%%%%%%%%%%%

\begin{figure}[htbp]
\begin{center}
\includegraphics[scale=1.2]{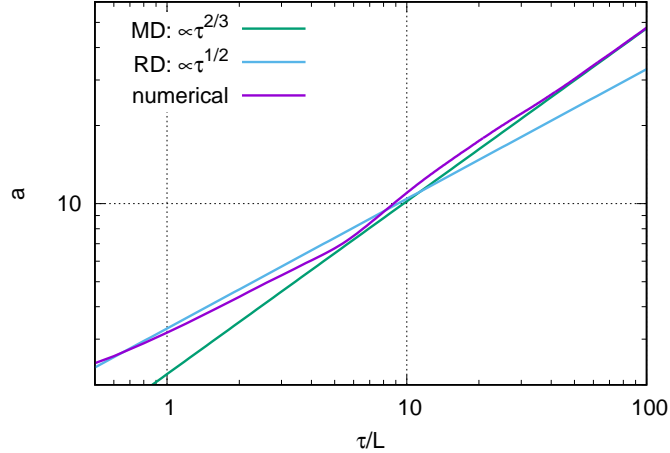}
\caption{The scale factor $a$ as a function of the proper time $\tau$ for $A=10$. 
}
\label{fig:atau10}
\end{center}
\end{figure}

Since the black hole is formed, one would expect that the expansion law of the lattice universe should approach to 
that of the matter dominated universe. 
Actually, as is shown in Fig.~\ref{fig:atau10}, the scale factor asymptotically approaches to 
the behavior $\propto \tau^{2/3}$ in contrast to Fig.~\ref{fig:atau1} for $A=0.1$. 
%%%%%%%%%%%%%%%%%%%%%%%%%%%<<start figure>>%%%%%%%%%%%%%%%%%%%%%%%%%%
%%%%%%%%%%%%%%%%%%%%%%%%%%%%<<end figure>>%%%%%%%%%%%%%%%%%%%%%%%%%%%
This result shows that the final fate of the gravitational collapse 
significantly affects the late time behavior of the lattice universe. 

\begin{figure}[htbp]
\begin{center}
\includegraphics[scale=1.2]{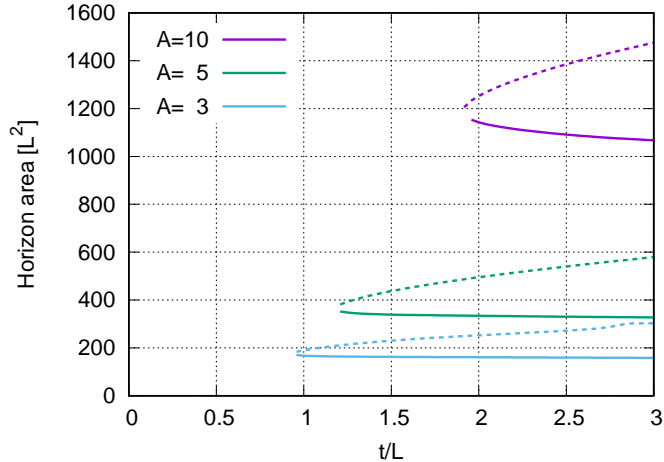}
\caption{Time evolution of the area of the 
horizon is shown for each value of the initial 
scalar field amplitude $A$. 
The dashed line shows the area of the inner intersection sphere 
between the time slice and the trapping horizon, 
and the solid line shows the area of the outer intersection sphere. 
}
\label{fig:area_Adep}
\end{center}
\end{figure}

Finally, let us check the time evolution of the trapping horizon in this model. 
A trapping horizon is a spacelike hyper-surface on which the outgoing null expansion vanishes. 
It comes to be observed when it intersects with a time slice of our numerical evolution. 
The intersection is given by two disconnected spheres on a time slice. 
In Fig.~\ref{fig:area_Adep}, the time evolution of 
the area for each disconnected piece is shown. 
The dashed line shows the area of the inner sphere and the solid line shows the area of the outer sphere. 

The horizon formation time and the area depends on the initial amplitude of the scalar field as 
is shown in Fig.~\ref{fig:area_Adep}. 
%%%%%%%%%%%%%%%%%%%%%%%%%%%<<start figure>>%%%%%%%%%%%%%%%%%%%%%%%%%%
%%%%%%%%%%%%%%%%%%%%%%%%%%%%<<end figure>>%%%%%%%%%%%%%%%%%%%%%%%%%%%
Since the initial scale of the inhomogeneity compared with the horizon scale 
$1/H$ is larger for a larger value of $A$
as is shown in Fig.~\ref{fig:AHi}, 
the Hubble horizon entry time of the scalar field bump is delayed for a larger value of $A$. 
As a result, the horizon formation time is also delayed and the horizon area 
at the formation time is larger for a larger value of $A$ as is explicitly shown in Fig.~\ref{fig:area_Adep}. 

\begin{figure}[htbp]
\begin{center}
\includegraphics[scale=1]{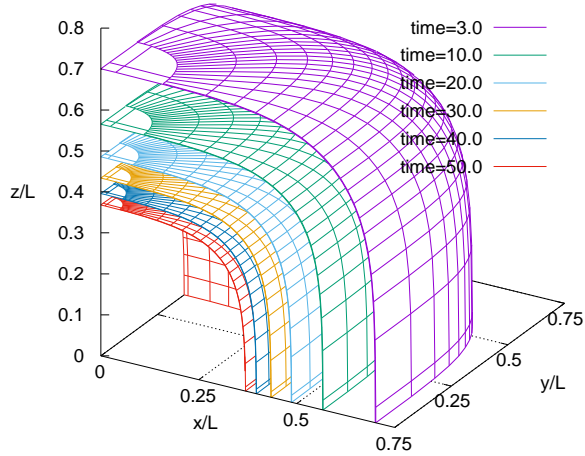}
\caption{Time evolution of the horizon shape for the $A=10$ case . 
}
\label{fig:horizon}
\end{center}
\end{figure}

\begin{figure}[htbp]
\begin{center}
\includegraphics[scale=1.2]{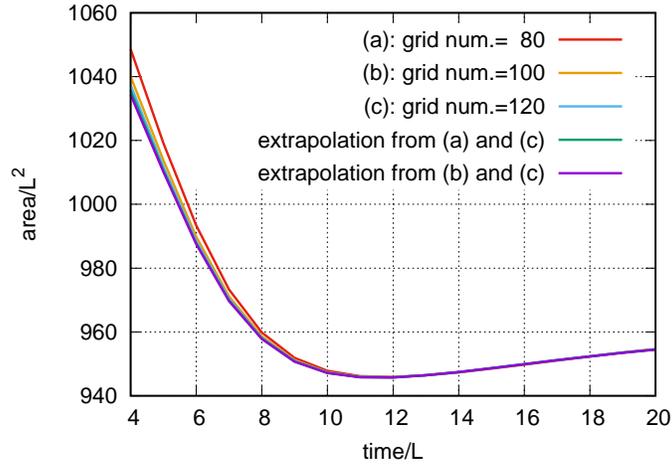}
\caption{Time evolution of the horizon area. 
The extrapolation is performed by assuming the quartic convergence. 
}
\label{fig:AH_evo}
\end{center}
\end{figure}

\begin{figure}[htbp]
\begin{center}
\includegraphics[scale=1.2]{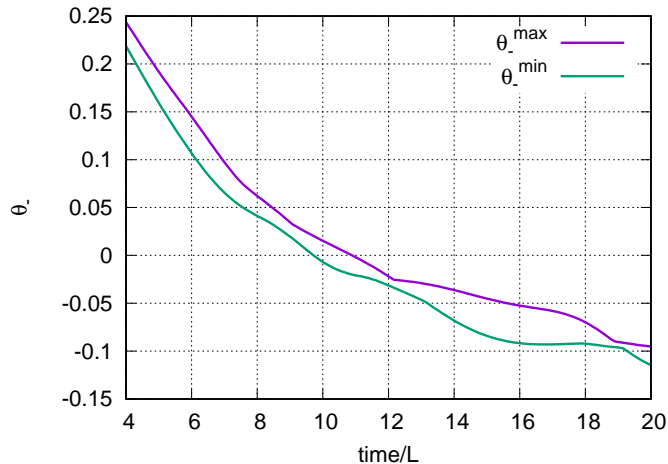}
\caption{The value of $\theta_-^{\rm max}$ and $\theta_-^{\rm min}$ on the marginally trapped surface as 
functions of the coordinate time. 
}
\label{fig:expansion}
\end{center}
\end{figure}

Hereafter, we focus on only the outer intersection sphere and refer to 
it as just horizon or trapping horizon. 
As is shown in Fig.~\ref{fig:horizon}, the coordinate radius of the 
horizon monotonically shrinks with time. 
%%%%%%%%%%%%%%%%%%%%%%%%%%%<<start figure>>%%%%%%%%%%%%%%%%%%%%%%%%%%
%%%%%%%%%%%%%%%%%%%%%%%%%%%%<<end figure>>%%%%%%%%%%%%%%%%%%%%%%%%%%%
On the other hand, the proper area of the horizon is not monotonic as is shown in Fig.~\ref{fig:AH_evo}. 
%%%%%%%%%%%%%%%%%%%%%%%%%%%<<start figure>>%%%%%%%%%%%%%%%%%%%%%%%%%%
%%%%%%%%%%%%%%%%%%%%%%%%%%%%<<end figure>>%%%%%%%%%%%%%%%%%%%%%%%%%%%
The horizon area is decreasing with time in the beginning. 
This behavior is not common in asymptotically flat cases. 
However, as is reported in Ref.~\cite{Harada:2005sc} for a spherically symmetric system, 
such decreasing area of a horizon can be realized by transition from a past outer trapping horizon(POTH) 
to a future outer trapping horizon(FOTH) defined in Ref.~\cite{Hayward:1993wb}. 
Let us define the orthonormal tetrad bases $n^\mu$, $s^\mu$, $\ee^\mu_{(1)}$ and $\ee^\mu_{(2)}$ 
on a marginally trapped surface, 
where $n^\mu$ is the unit normal vector to a time slice introduced in Sec.~\ref{sec-2}, 
and $s^\mu$ is normal to the marginally trapped surface, so that $\ee^\mu_{(1)}$ and $\ee^\mu_{(2)}$
are tangent to the marginally trapped surface. 
Then, the expansions $\theta_\pm$ associated with two null vectors $k^\mu_\pm:=(n^\mu\pm s^\mu)/\sqrt{2}$ can be expressed as 
\begin{equation}
\theta_\pm:=\pm D_i s^i + K_{ij} s^i s^j -K. 
\end{equation}
The marginally trapped surface on FOTH(POTH) satisfies the following condition:
\begin{eqnarray}
\theta_+&=&0~{\rm and}~\theta_-<0~{\rm for~FOTH},\\
\theta_+&=&0~{\rm and}~\theta_->0~{\rm for~POTH}.
\end{eqnarray}
We note that the value of $\theta_-$ is not constant on a marginally trapped surface 
for a given time slice, 
and is not necessarily positive or negative definite. 
Therefore, we call it FOTH(POTH) when $\theta_-<0(>0)$ everywhere on the marginally trapped surface. 
From the theorem in Ref.~\cite{Hayward:1993wb}, 
the area of FOTH(POTH) should increase(decrease) with the parameter associated with 
a horizon generator $l^\mu$ whose orientation is fixed by $k_+^\mu l_\mu>0$. 
For the portion of the trapping horizon which we are focusing on, the parameter associated with $l^\mu$ 
is an increasing function of time. 
Therefore, the area of FOTH(POTH) should increase(decrease) with time. 
We plot the maximum value $\theta_-^{\rm max}$ and the minimum value $\theta_-^{\rm min}$ of 
$\theta_-$ on the marginally trapped surface as functions of time 
in Fig.~\ref{fig:expansion}. 
%%%%%%%%%%%%%%%%%%%%%%%%%%%<<start figure>>%%%%%%%%%%%%%%%%%%%%%%%%%%
%%%%%%%%%%%%%%%%%%%%%%%%%%%%<<end figure>>%%%%%%%%%%%%%%%%%%%%%%%%%%%
As is shown in Fig.~\ref{fig:expansion}, the value of $\theta_-$ changes its sign 
around $t\sim11L$, which is similar to the time when the area of the marginally trapped surface takes 
the minimum value in Fig.~\ref{fig:AH_evo}. 
Therefore we conclude that 
the time evolution of the horizon area is compatible with 
the area theorem given in Ref.~\cite{Hayward:1993wb} within the accuracy of our numerical simulation.%
\footnote{
\baselineskip5mm
If we look at the relation between the sign of $\theta_-$ and the time evolution of the horizon area more precisely, 
we find that the horizon area is still decreasing at the moment $\theta_-^{\rm max}=0$. 
However, since the difference between the area at the moment  $\theta_-^{\rm max}=0$ and the minimum value of the area 
is $0.01\%$, we would be able to conclude that 
the time evolution of the horizon area is compatible with 
the area theorem within the accuracy of our numerical simulation. 
}
 
\begin{figure}[htbp]
\begin{center}
\includegraphics[scale=1]{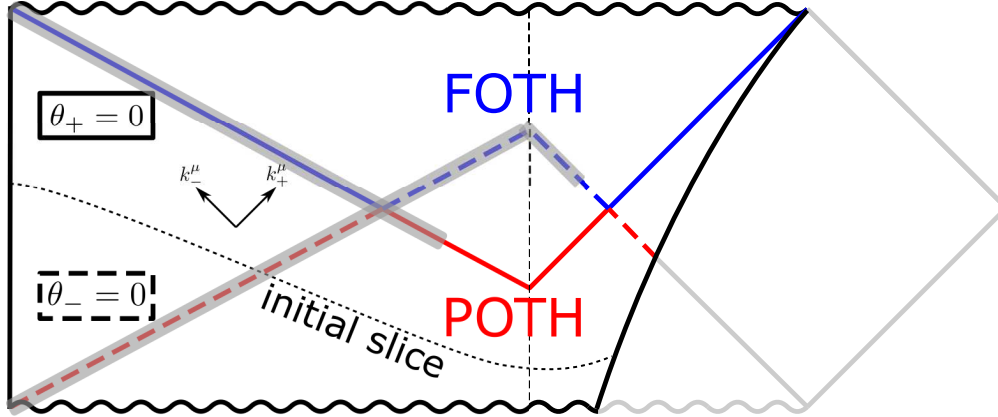}
\caption{A schematic figure for the spacetime structure. 
The blue and red lines denote FOTH and POTH, respectively. 
The solid and the dashed lines denote the horizons with $\theta_+=0$ and $\theta_-=0$, respectively. 
We could not confirm the horizon structure in the shaded part for our model because of the 
limitation of the resolution. 
}
\label{fig:schematic}
\end{center}
\end{figure}

\begin{figure}[htbp]
\begin{center}
\includegraphics[scale=1.2]{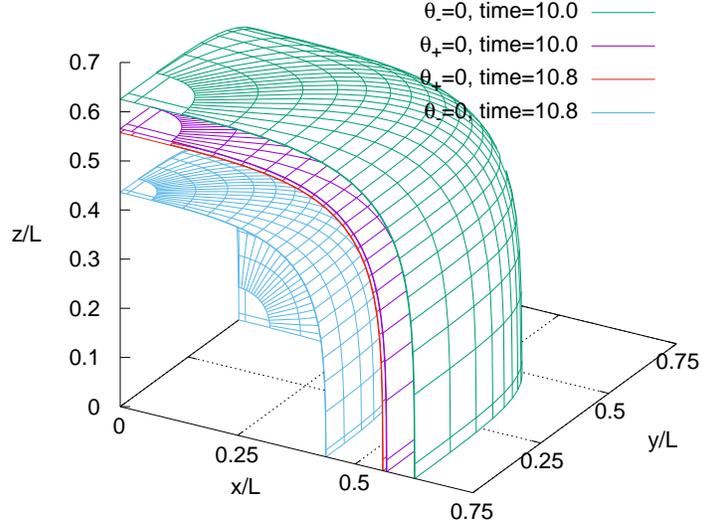}
\caption{
Snapshots of the trapping horizons with $\theta_+=0$ and $\theta_-=0$ are shown at 
$t=10L$ and $t=10.8L$. 
The surface with $\theta_-=0$ is initially outside the surface $\theta_+=0$, 
and shrinks inside the surface with $\theta_+=0$ at $t=10.8L$. 
The region bounded by the outer green surface($\theta_-=0$) and the inner purple surface($\theta_+=0$) 
is past-trapped at $t=10L$. 
While, the region bounded by the outer red surface($\theta_+=0$) and the inner blue surface($\theta_-=0$) 
is future-trapped at $t=10.8L$. 
}
\label{fig:passthrough}
\end{center}
\end{figure}

The spacetime structure can be schematically described by jointing a part of a closed homogeneous universe model 
and a part of the Schwarzschild spacetime as is shown in Fig.~\ref{fig:schematic}. 
%%%%%%%%%%%%%%%%%%%%%%%%%%%<<start figure>>%%%%%%%%%%%%%%%%%%%%%%%%%%
%%%%%%%%%%%%%%%%%%%%%%%%%%%%<<end figure>>%%%%%%%%%%%%%%%%%%%%%%%%%%%
Then, one may expect that, around the transition from POTH to FOTH, 
the trapping horizon passes through another trapping horizon associated with the surface on which 
$\theta_-=0$ is satisfied. 
Actually, we observe this behavior finding the evolution of the trapping horizon with $\theta_-=0$. 
\footnote{
In Ref.~\cite{Hayward:1993wb}, differently from our definition, 
the subscript $+$ is always assigned to the direction 
for which the expansion vanishes.
}
In Fig.~\ref{fig:passthrough}, we show the snapshots of the trapping horizons with $\theta_+=0$ and $\theta_-=0$ at 
$t=10L$ and $t=10.8L$. 
The figure clearly shows the surface $\theta_-=0$ passes through the surface $\theta_+=0$ and 
shrinks inside it. 
The shell region 
bounded by the two trapping horizons corresponds to either a black hole($\theta_+=0$ on the outer boundary) 
or a white hole region($\theta_+=0$ on the inner boundary). 
%whose outer bounary and inner boundary are respectively given by $\theta_-=0$($\theta_+=0$) and $\theta_+=0$($\theta_-=%0$) is past-trapped(future-trapped) and corresponds to a white hole(black hole) region. 
%%%%%%%%%%%%%%%%%%%%%%%%%%%<<start figure>>%%%%%%%%%%%%%%%%%%%%%%%%%%
%%%%%%%%%%%%%%%%%%%%%%%%%%%%<<end figure>>%%%%%%%%%%%%%%%%%%%%%%%%%%%
We could not confirm the horizon structure in the shaded region in Fig.~\ref{fig:schematic}
for our model because of the limitation of the resolution.

%%%%%%%%%%%%%%%%%%%%%%%%%%%%%%%%%%%%%%%%%%%%%%%%%%%%%%%%%%%%%%%%
\section{summary and discussion}
%%%%%%%%%%%%%%%%%%%%%%%%%%%%%%%%%%%%%%%%%%%%%%%%%%%%%%%%%%%%%%%%

We have investigated gravitational collapse of a massless scalar field 
with the periodic boundary condition in a cubic box. 
This system can be regarded as a lattice universe model and the expansion law of 
the lattice universe has been also checked. 
We have shown that there are two typical behaviors 
depending on the initial amplitude of the scalar field bump. 
For a relatively small amplitude, the scalar field diffuses into whole box region and 
continues to oscillate. 
The amplitude of the oscillation decays with time and the expansion law of the lattice universe 
approaches to that of the radiation dominated universe. 
In contrast to the small amplitude case, 
for a large enough amplitude, a black hole is formed as a result of the gravitational collapse of 
the scalar field bump. 
The expansion law of the lattice universe approaches to that of the matter dominated universe. 
This result clearly shows that the late time expansion of the lattice universe 
can be significantly affected by the fate of the gravitational collapse. 
We found that, in the initial stage, the horizon is classified into the past outer 
trapping horizon~\cite{Hayward:1993wb} 
and the area of the horizon decreases with time. 
After a period of time, the transition from the past outer trapping horizon to 
the future outer trapping horizon can be observed and the horizon area starts to increase. 

In this paper, we could not reveal the details around the critical amplitude which 
divides the phase of the gravitational collapse and the asymptotic behavior of the lattice universe. 
The difficulty arises from the resolution problem. 
Since the coordinate radius of the black hole horizon shrinks with time, 
it is harder to resolve the horizon for a long time for 
a smaller initial amplitude. 
The scale-up coordinate introduced in Sec.~\ref{sec-2} helps 
to resolve it to some extent. 
However, since the scale-up coordinate makes the resolution near the boundary worse, 
the parameter for the scale-up coordinate also has a limitation. 
Furthermore, it is well known that, 
in order to see the critical behavior\cite{Choptuik:1992jv} of the gravitational collapse, 
extremely high resolution is needed even for asymptotically flat cases. 
Nevertheless, the two examples shown in this paper are enough to 
understand the general picture of the gravitational collapse with a periodic boundary condition. 
Some features in this system would be common in gravitational collapse 
in a cosmological background as in the case of primordial black hole formation\cite{1967SvA....10..602Z,Hawking:1971ei}.

%%%%%%%%%%%%%%%%%%%%%%%%%%%%%%%%%%%%%%%%%%%%%%%%%%%%%%%%%%%%%%%%
\section*{Acknowledgements}
%%%%%%%%%%%%%%%%%%%%%%%%%%%%%%%%%%%%%%%%%%%%%%%%%%%%%%%%%%%%%%%%
We thank V. Cardoso and T. Harada for helpful comments.  
The authors thank the Yukawa Institute for Theoretical Physics at Kyoto University. 
Discussions during the YITP workshop YITP-T-18-05 on "Dynamics in Strong
Gravity Universe" were useful to complete this work. 
T.I. acknowledges financial support provided under the European Union's H2020 ERC Consolidator Grant gMatter and strong-field gravity: New frontiers in Einstein's theoryh grant agreement no. MaGRaTh-646597, and under the H2020-MSCA-RISE-2015 Grant No. StronGrHEP-690904. 
This work was supported by JSPS KAKENHI Grant
Number JP16K17688 (C.Y.).

%\appendix

%\bibliographystyle{h-physrev5-title}
%\bibliography{../../bibfiles/swcollapse}

\end{document}